\documentclass[aps,prl,reprint,amsmath,amssymb,superscriptaddress, 10pt]{revtex4-2}
\usepackage{graphicx}
\usepackage[utf8]{inputenc}
\usepackage{amsmath}
\usepackage{subfigure}
\usepackage{pdfpages}
\makeatletter
\AtBeginDocument{\let\LS@rot\@undefined}
\makeatother
\usepackage{amsfonts}
\usepackage{amssymb}
\usepackage{bm}
\usepackage{dcolumn}
\usepackage{array}  
\usepackage{diagbox}
\usepackage{color}
\usepackage[utf8]{inputenc}
\usepackage[T1]{fontenc}
\usepackage{booktabs, array, mathptmx, float, tabularx, booktabs, lipsum, amsmath, multirow,mathtools}
\graphicspath{{figs/}{figsgaoerb/}}
\usepackage{soul}
\usepackage{xcolor}
\usepackage{mathtools}
\usepackage{lipsum}
\usepackage{comment}

\usepackage{xcolor}
\usepackage[colorlinks,citecolor=blue,linkcolor=blue,anchorcolor=blue,urlcolor=blue]{hyperref}

\begin{document}

\title{Finite-time Polytropic Steering Reveals Non-monotonic Irreversibility}

\author{Cong Fu}
\affiliation{School of Physics and Astronomy, Beijing Normal University, Beijing 100875, China}
\affiliation{Department of Physics, Xiamen University, Xiamen 361005, Fujian, China}

\author{Youhui Lin}
\affiliation{Department of Physics, Xiamen University, Xiamen 361005, Fujian, China}

\author{Shanhe Su} \email{sushanhe@xmu.edu.cn}
\affiliation{Department of Physics, Xiamen University, Xiamen 361005, Fujian, China}

\author{Yu-Han Ma}
\email{yhma@bnu.edu.cn}
\affiliation{School of Physics and Astronomy, Beijing Normal University, Beijing 100875, China}
\affiliation{Key Laboratory of Multiscale Spin Physics (Ministry of Education), Beijing Normal University, Beijing 100875, China}

\begin{abstract}

  How irreversibility evolves from slow to fast driving is a central question in nonequilibrium thermodynamics. While near-equilibrium theory predicts increasing irreversibility with driving speed, the corresponding crossover far from equilibrium is not generally resolved within a unified control framework. Here we introduce polytropic steering, in which a prescribed relation between the ensemble energy and a time-dependent control parameter continuously organizes finite-time processes between isothermal and adiabatic limits. A Fokker-Planck formulation makes the construction applicable to general confining potentials, and yields closed-form thermodynamic expressions for a harmonic trap. We find a non-monotonic irreversibility: the entropy generation decreases in both the quasistatic and fast-driving limits and exhibits a maximum at an intermediate protocol duration. The resulting most-irreversible timescale quantifies the competition between thermal relaxation and external driving. We verify the persistence of this behavior in an anharmonic optical-lattice potential and demonstrate how the same steering principle controls power-efficiency trade-offs in finite-time Brownian engines.
\end{abstract}

\maketitle

\textit{Introduction.---}Finite-time control of stochastic systems is a central problem in nonequilibrium
statistical physics, with broad implications for 
thermal machines, information processing, and driven mesoscopic matter~\cite{esposito2009,campisi2011,pekola2015,vinjanampathy2016,seifert2012,blickle2012,quinto-su2014,martinez2016,proesmans2020,chen2022,zhao2024}. The classical Carnot cycle~\cite{Carnot1872} provides the ideal benchmark of reversible operation, but its quasistatic isothermal and adiabatic branches require infinitely slow evolution. Finite-time operation therefore creates a fundamental trade-off between
operational speed and thermodynamic cost, making the dependence of irreversibility on driving time essential to the characterization and control of finite-time thermodynamic processes~\cite{salamon1980,jarzynski2011,kato2020,brunelli2018,landi2021,seara2021,roduner2022}. In the near-equilibrium regime, irreversible entropy generation (IEG) decreases with protocol duration ($\tau$) as $1/\tau$~\cite{salamon1983,Sekimoto1997,schmiedl2007,berut2012,cavina2017,ma2018,ma2020,vanvu2021}. This scaling underlies a broad class of finite-time thermodynamic theories and optimization principles for thermodynamic cycles~\cite{rubin1979,salamon1980,esposito2010,maUniversal2018,yuan2022,ma2025}.

The $1/\tau$-scaling, however, does not generally extend to the fast-driving regime. When the protocol timescale approaches or falls below the characteristic relaxation time of the system in its thermal environment, the exchanged work and heat can be suppressed, leading to deviations from the conventional $1/\tau$ behavior and, in some settings, to reduced dissipation at short times ~\cite{deng2013,deffnerKibbleZurek2017,villazon2019,ma2020,cavina2021,ma2021}. Such short-time deviations can have important consequences for finite-time thermodynamic optimization, including the validity of low-dissipation description of heat engines~\cite{ma2021,cavina2021}. Related suppression mechanisms have also been reported in rapidly driven quantum systems~\cite{claeys2019,villazon2019,pekola2019}.
These results indicate that the usual relation between driving speed and dissipation can change qualitatively away from the near-equilibrium regime. Yet a unified description of irreversibility across the quasistatic and fast-driving regimes is still lacking. In particular, it is unclear whether the crossover between them gives rise to a characteristic timescale of maximal irreversibility.

Here we address this question by introducing \textit{finite-time polytropic steering} for Brownian particles in a time-dependent confining potential. The approach prescribes a thermodynamic relation between the ensemble energy and the externally controlled potential parameter, and combines this constraint with the underlying Fokker-Planck dynamics and ensemble law of thermodynamics. This yields a general steering equation applicable to confining potentials, while the polytropic index serves as a control parameter that continuously connects the quasistatic isothermal and
adiabatic limits. The resulting framework provides a unified description of finite-time Brownian dynamics across the slow-to-fast crossover, while retaining the nonequilibrium evolution of the probability distribution.

\textit{Polytropic steering of Brownian particles}.--- We consider an overdamped Brownian particle confined by a time-dependent potential $U(x,\lambda(t))$, where $\lambda(t)$ is an externally controlled parameter that modulates the potential landscape. The particle is coupled to a thermal reservoir at temperature $T_{\mathrm{s}}$ and subject to viscous damping with coefficient $\gamma$. Its probability density $\rho(x,t)$ evolves according to the overdamped FP equation~\cite{Edward1967,Holubec2019,shiraishi2023,foot2}
\begin{equation}
	\frac{\partial \rho(x,t)}{\partial t} = \frac{\partial}{\partial x} \left[ \frac{1}{m\gamma} \left( \frac{\partial U}{\partial x} \rho(x,t) + T_s \frac{\partial \rho(x,t)}{\partial x} \right) \right],
	\label{Smoluchowski equation}
\end{equation}
where $m$ is the particle mass. The initial trap defines a characteristic spatial relaxation time $\tau_x \equiv m\gamma/\lambda(t=0)$, which sets the timescale for the particle to relax towards equilibrium. We set Boltzmann constant $k_B=1$ throughout this Letter. We define a finite-time polytropic process through the invariant relation
\begin{equation}
	E (t) \lambda^\xi(t)=\mathrm{const},\quad \xi \in \mathbb{R}.
	\label{eq:polytropic process E}
\end{equation}
Here, $E(t) \equiv \langle U(x, \lambda(t)) \rangle$, and the ensemble average with respect to the probability density $\rho(x,t)$ is defined as
$\langle \cdot \rangle \equiv \int_{-\infty}^{\infty} (\cdot)\,\rho(x,t)\,\mathrm{d}x$. This definition applies arbitrarily far from equilibrium. To derive the control protocol that enforces Eq.~\eqref{eq:polytropic process E}, we start from the first law of thermodynamics at the ensemble level,
\begin{equation}
	\dot E (t) =\dot \lambda  \langle \partial_\lambda U \rangle+\dot Q(t),
	\label{eq:1st law}
\end{equation}
where the first term on the right-hand side represents the instantaneous work rate, and $\dot Q \equiv \int_{-\infty}^{\infty} U(x, \lambda(t))\, \partial_t \rho(x, t) \,\mathrm{d}x$ denotes the heat flux~\cite{martinez2015,martinez2016,salazar2019,shiraishi2023}. Substituting Eq.~\eqref{Smoluchowski equation} yields the exact heat flux
\begin{equation}
\dot{Q}(t) = -\frac{1}{m\gamma} \Big[ \langle (\partial_x U)^2 \rangle - T_s \langle \partial_{xx} U \rangle \Big].
\label{eq:exact_Q_kin}
\end{equation}
Differentiating Eq.~\eqref{eq:polytropic process E} gives $\dot E/E=-\xi \dot \lambda / \lambda$. Equating this with Eq.~\eqref{eq:1st law} and substituting Eq.~\eqref{eq:exact_Q_kin}, we obtain the generalized polytropic steering equation
\begin{equation}
\dot{\lambda}(t) = \frac{ \lambda \Big[ \langle (\partial_x U)^2 \rangle - T_s \langle \partial_{xx} U \rangle \Big] }{m \gamma \Big[ \lambda \langle \partial_\lambda U \rangle + \xi E \Big] }.
\label{eq:over_general_proto}
\end{equation}
Equation~\eqref{eq:over_general_proto} is one of the central results of this Letter. For a specified polytropic index $\xi$ and potential $U(x,\lambda)$, the coupled Eqs.~\eqref{Smoluchowski equation} and \eqref{eq:over_general_proto} form a closed dynamical system that uniquely determines the protocol $\lambda(t)$ together with the full nonequilibrium evolution of $\rho(x,t)$. Equation~\eqref{eq:over_general_proto} thus provides an exact and experimentally implementable family of finite-time polytropic protocols, hereafter referred to as {\it{polytropic steering}}. 

While Eq.~\eqref{eq:over_general_proto} generally depends on $\rho(x,t)$ and must be solved with Eq.~\eqref{Smoluchowski equation} as a coupled integro-differential system, the description simplifies dramatically for the experimentally relevant case of a harmonic potential $U(x,t)=(k_t x^2)/2$ (with $k_t$ the trap stiffness). For clarity, we adopt this choice throughout the main text and the End Matter demonstrates that the universal features persist for anharmonic potentials. In the harmonic case, the Brownian system can be described by the effective temperature $\theta(t)$, since the potential energy regulation allows for instantaneous equilibrium distributions~\cite{foot1}. Starting from a state characterized by $\theta(t=0) \equiv \theta_{0}=\left( 1+\delta \right)T_{\rm{s}}$ (with $\delta$ measuring the initial temperature mismatch), the distribution remains exactly instantaneous  canonical form distribution, characterized by $\theta(t)\equiv 2\left \langle E(t)\right \rangle$ for $t \in [0,\tau]$, as obtained from the position variance. With this description, Eq.~\eqref{eq:over_general_proto} can be expressed as an implicit ordinary differential equation (ODE) control protocol for $k_t$
\begin{equation}
	\frac{\dot{k}_t}{k_t} = \frac{2k_t}{m \gamma (\xi + 1)} \left[ 1 - \frac{1}{1+\delta } \left(\frac{k_t}{k_0}\right)^\xi \right].
	\label{eq:poly control overdamped}
\end{equation}
Experimentally, such a state can be prepared by applying a Gaussian white-noise electric field, which sets the initial effective temperature. Concurrently, the trap stiffness is controlled by modulating the laser power with millisecond resolution~\cite{Martnez2013,martinez2016}.

\begin{figure}
\includegraphics[width=0.48\textwidth]{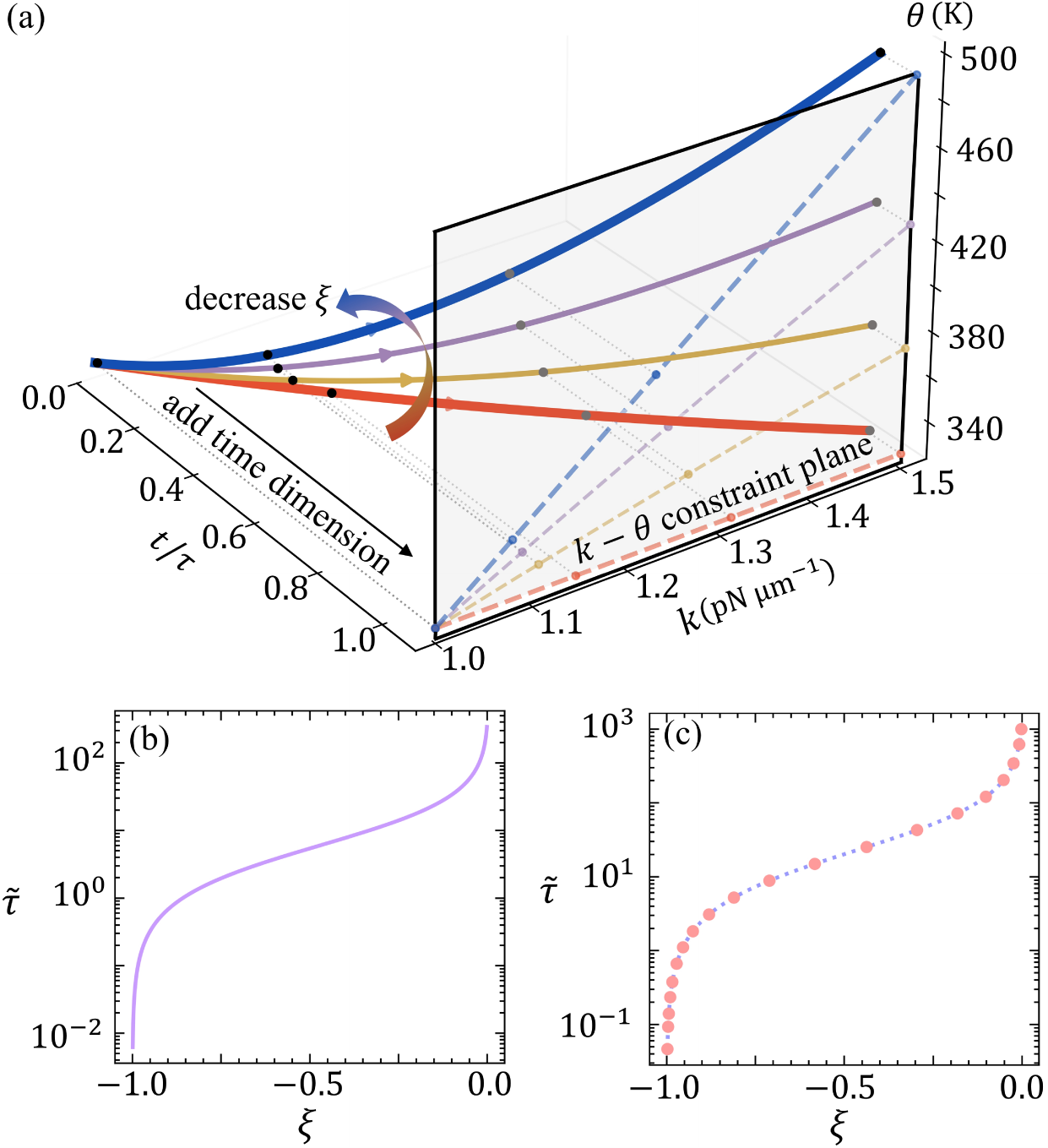}
	\caption{(a) The finite-time evolution of the thermodynamic state $(k,\theta)$ as a function of normalized time $t/\tau$ for polytropic index $\xi=0$ (red), $\xi=-0.33$ (yellow), $\xi=-0.66$ (purple), and $\xi=-1$ (blue). The colored arrows, graded from red to blue, mark the decrease of the polytropic index $\xi$ from the isothermal to the adiabatic limit. The dashed curves projected on the shaded gray plane are given by the polytropic constraint $\theta\lambda^{\xi}={\rm{const.}}$ The dimensionless process duration $\tilde{\tau} \equiv \tau/\tau_x$ as a function of $\xi$ for underdamped regime (b) and overdamped regime (c), where simulation data are represented by pink dots. In this figure, $u=1.5$ and $\delta=1 \times 10^{-3}$ are used.}
	\label{fig:poly_control}
\end{figure}

Traditionally, polytropic processes are introduced exclusively for an ideal gas undergoing a quasistatic succession of equilibrium states, which collapses the temporal dimension~\cite{callen1985}. Polytropic steering extends this conventional framework by explicitly incorporating the time dimension, as illustrated in the spatiotemporal diagram [Fig.~\ref{fig:poly_control}(a)]. Our approach reveals that the finite-time path is uniquely determined by the coupling between the prescribed energy evolution [Eq.~\eqref{Smoluchowski equation}] and the underlying geometric constraints [Eq.~\eqref{eq:polytropic process E}] of the parametric space. As a novel illustration of the overdamped design, for $\xi=0$, Eq.~\eqref{eq:poly control overdamped} yields $k^{-1}_t=k_0^{-1}-2\delta t/[m \gamma(1+\delta)]$, furnishing an analytically tractable protocol for a quasi‑isothermal stroke (the system remains isothermal but at a temperature different from the bath temperature) and thus providing the isothermal branch of an endoreversible cycle~\cite{curzon1975,chen2022}. A feature of the proposed polytropic steering paradigm is its independence from the specific damping regime. While the framework is initially developed for the overdamped case, it extends naturally to the underdamped regime by replacing the underdamped FP equation~\cite{foot1}. In this regime, the protocol translates from an implicit ODE into an explicit protocol for the harmonic potential as 
\begin{equation}
	\beta(t)  = \beta_{0} \left[ \frac{\delta}{1+\delta}\exp \left (- \frac{\xi \gamma}{\xi+1}t \right ) +\frac{1}{1+\delta} \right ]^{-\frac{1}{\xi}},
	\label{eq:poly control underdamped}
\end{equation}
where $\beta_0 \equiv \beta(t=0) \equiv k_0^{1/2}$. Note that polytropic steering requires $\delta \neq 0$ initial states; otherwise, only isochoric (constant stiffness) or isothermal processes are possible. Equations~\eqref{eq:poly control overdamped} and \eqref{eq:poly control underdamped} further show that the initial effective temperature relative to the reservoir controls the process direction: $\delta < 0$ for expansion (decrease in stiffness), while $\delta > 0$ for compression (increase in stiffness). A finite temperature offset is not merely an initial condition but a prerequisite for finite-time polytropic steering~\cite{foot1}: at $\delta=0$, the system reduces to the reversible isothermal limit, where no finite-time control along a polytropic trajectory can be initiated. Particularly, for the quasi-isothermal process of $\xi=0$, Eq.~\eqref{eq:poly control underdamped} reduces to $\beta(t)=\beta_0 \left( \beta_{\mathrm{f}} /\beta_0\right)^{t/\tau}$ with $\beta(\tau)\equiv\beta_{\rm{f}}$, which recovers the result of Ref.~\cite{chen2022}. The key to achieving polytropic steering with different $\xi$ is to change the control duration of the process. It follows from Eq.~\eqref{eq:poly control underdamped} that
\begin{equation}
	\tau=-\gamma^{-1}(1+\xi^{-1})\ln \left[ (1+\delta^{-1})u^{-\xi}-\delta^{-1}
	\right],
	\label{eq:tau_xi_norm}
\end{equation}
where $u \equiv \beta_{\mathrm{f}} /\beta_0 $. As shown in Fig.~\ref{fig:poly_control}(b), achieving $\xi$ from $0$ (isothermal) to $-1$ (adiabatic) requires tuning driving speed from slow to fast. 
\begin{figure}
	\centering
\includegraphics[width=0.48\textwidth]{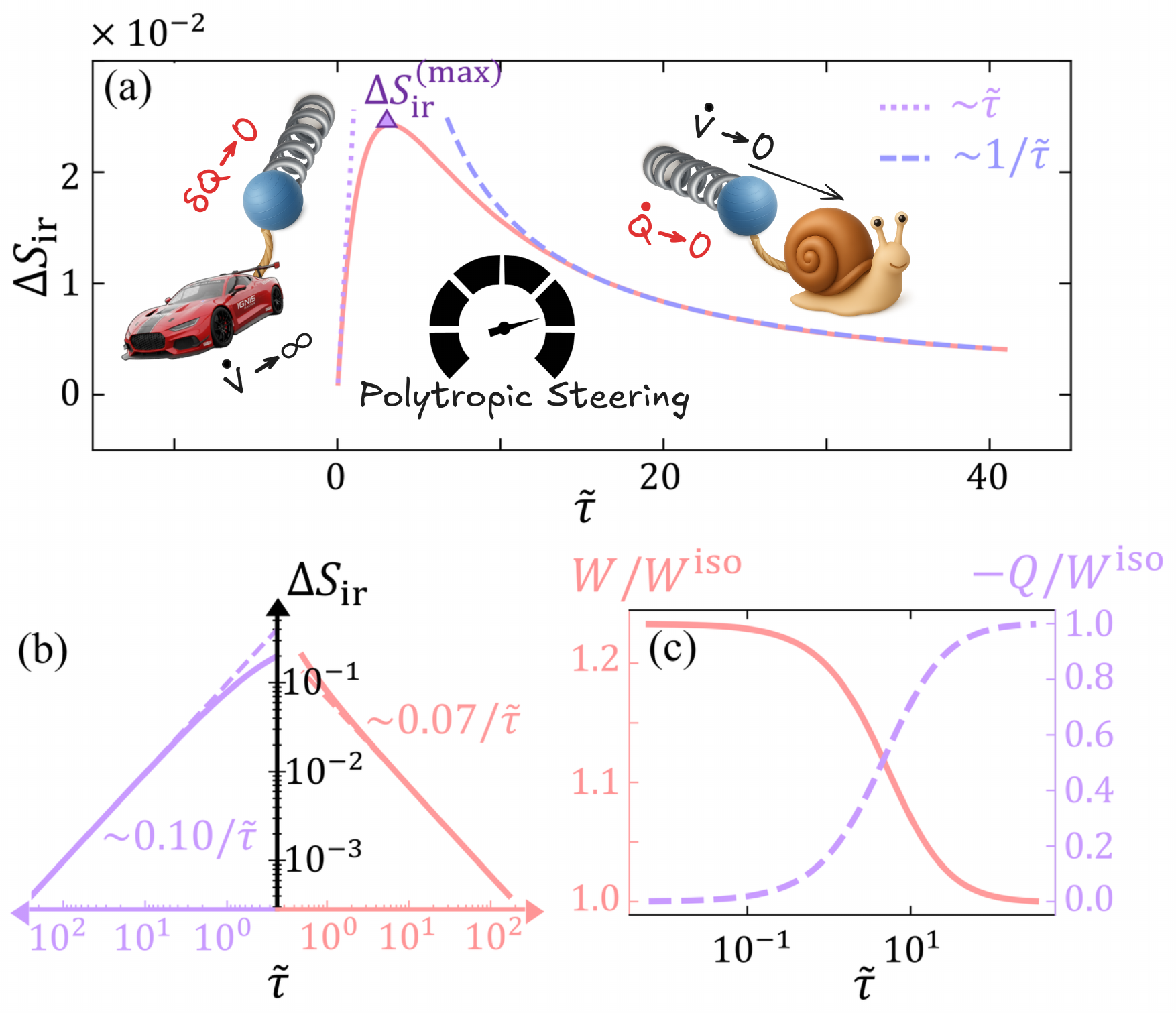}
	\caption{(a) IEG $\Delta S_{\mathrm{ir}}$ as a function of the dimensionless process duration $\tilde{\tau}\equiv \tau/\tau_x$ (solid red curve). (b) $\Delta S_{\mathrm{ir}}$ as a function of $\tilde{\tau}$ for quasistatic isothermal processes (solid curve, $\xi=0$). The left (right) axis is associated with with $u=3/2$ ($u=2/3$). The dashed line represent the predicted asymptotic scaling. (c) Normalized work $W$ (pink solid curve) and heat $-Q$ (purple dashed curve) as a function of $\tilde{\tau}$, where $W_{\mathrm{iso}}=\theta_0 \ln u$. To plot (a) and (c), we set $u=1.5$ and $\delta=1 \times 10^{-2}$.}
	\label{fig:S_energy}
\end{figure}
Despite the implicit nature of the overdamped protocol, the resulting $\tilde{\tau}$-$\xi$ relation exhibits the same qualitative behavior as the underdamped case: $\tilde{\tau}$ increases monotonically from $0$ at $\xi=-1$ to $\infty$ at $\xi\to 0$, as validated by our numerical solutions illustrated in Fig.~\ref{fig:poly_control}(c). The simulation data (pink dots) validate the theoretical result (blue dotted curve) of Eq.~\eqref{eq:poly control overdamped}. 

Note that the required timescale separation, which depends on the level of description employed and is of central importance in stochastic thermodynamics~\cite{Edward1967,salazar2019}, underpins the above finite-time control task~\cite{foot1}. For a general potential, the present framework is formulated directly at the FP level and therefore requires only the standard conditions underlying stochastic dynamics: $\tau \gg \tau_p \equiv 1/\gamma$ for overdamped dynamics after adiabatic elimination of momentum~\cite{Edward1967,seifert2012,shiraishi2023}, and $\tau \gg \tau_c$ for underdamped dynamics in the white-noise limit, where $\tau_c$ denotes the bath correlation time (see Supplemental Material for details under harmonic potential). These conditions impose no restriction on the relation between the protocol duration $\tau$ and the spatial relaxation time $\tau_x$, thereby covering both slow ($\tau \gg \tau_x$) and fast ($\tau \ll \tau_x$) driving regimes. These two regimes correspond to the isothermal limit $\xi=0$ (slow-driving, $\tilde{\tau}\equiv \tau/\tau_x \gg 1$) and the adiabatic limit $\xi=-1$ (fast-driving, $\tilde{\tau} \ll 1$), respectively. Furthermore, to illustrate the universality of this framework, we also numerically solve the coupled system in the End Matter for an optical lattice potential.

\textit{Irreversibility and energetics of polytropic steering}.--- Recent experimental \cite{ma2020} and numerical \cite{deffnerKibbleZurek2017} findings show that IEG in the fast-driving regimes departs from the typical $1/\tau$-scaling. As the driving time increases, the change of IEG as it varies across the short- to long-time regime has not yet been revealed. In the following, we reconcile the IEG behaviors quantitatively across different time regimes using the framework of finite-time polytropic processes. 

We begin with the general overdamped dynamics. The total IEG rate is the sum of the reservoir entropy generation rate and the system entropy generation rate $\dot \Sigma(t)=\dot S_{\mathrm{r}}(t)+\dot S_{\mathrm{s}}(t)$. Here and throughout, this IEG is evaluated for the working substance and its surrounding reservoir only, excluding any dissipative cost associated with the external driving apparatus itself. The entropy change rate of the reservoir at temperature $T_{s}$ is $\dot S_{\mathrm{r}}(t)=-\dot Q(t)/T_{s}$, where $Q(t)$ is given by Eq.~\eqref{eq:exact_Q_kin}. The system entropy is quantified by the Shannon entropy $S_{\mathrm{s}}=-\int_{-\infty}^{\infty}\rho(x,t) \ln \rho(x,t) dx$~\cite{seifert2005}. Substituting Eqs.~\eqref{Smoluchowski equation} and \eqref{eq:exact_Q_kin}, the total IEG rate for an confining potential reads
\begin{equation}
    \dot \Sigma(t)= \frac{1}{m\gamma T_s}  \langle (\partial_x U)^2 \rangle -\frac{2}{m\gamma } \langle \partial_{xx} U \rangle + \frac{T_s}{m\gamma} \mathcal{I}_F(t),
    \label{general_IEG_over_2}
\end{equation}
where $\mathcal{I}_F(t)\equiv \int_{-\infty}^{\infty} (\partial_x \rho)^2/\rho \,dx$ denotes the Fisher information~\cite{ito2020,melo2025}. Equation~\eqref{general_IEG_over_2} is the central result for the thermodynamic cost in the general framework. For a generic anharmonic potential, the IEG $\Delta S_{\mathrm{ir}}=\int_{0}^{\tau} \dot \Sigma(t) \, dt$ depends on the full time-dependent distribution $\rho(x,t)$ and must be evaluated numerically alongside the coupled steering dynamics (an example for the optical lattice potential is given in the End Matter). Remarkably, for the harmonic potential, Eq.~\eqref{general_IEG_over_2} can be integrated analytically as
\begin{equation}
	\Delta S_{\mathrm{ir}}= \chi (\xi+1)\left[\xi^{-1}(1+\delta)(1-u^{-\xi})-\mathrm{ln}u\right],
	\label{eq:normalized irrevisible entropy}
\end{equation}
where $\chi = 1 $~\cite{foot1} for underdamped; while $\chi=1/2$ for overdamped.

The dependence of the IEG on the protocol duration $\tilde{\tau}$ is shown in Fig.~\ref{fig:S_energy}(a), obtained by tuning $\xi$ [via Eqs.~\eqref{eq:tau_xi_norm} and \eqref{eq:normalized irrevisible entropy}]. Two asymptotic scalings emerge: in the fast‑driving limit ($\tilde{\tau} \ll 1$), the IEG grows linearly with duration; in the quasistatic limit ($\tilde{\tau} \gg 1$), it follows the universal $1/\tilde{\tau}$ decay. Note that the short time discussed above is meaningful only within the time-scale separation discussed in the polytropic steering before. The factors of these scalings depend on $\delta$ and $u$ are provided in the Supplemental Material~\cite{foot1}. At intermediate $\tilde{\tau}$ the IEG is nonmonotonic and reaches a maximum (purple triangle in Fig.~\ref{fig:S_energy}(a)), which defines a most‑irreversible protocol. This peak reflects the competition between the instantaneous dissipation rate and the available process time, analogous to observations in quantum thermodynamic systems~\cite{wang2021}. Although the closed-form expression in Eq.~\eqref{eq:normalized irrevisible entropy} is derived for an harmonic trap, its features, such as the non-monotonic peak and the parameter dependence, persist beyond Gaussian dynamics, as demonstrated by the optical lattice potential (see End Matter).

Additionally, one may also vary $\delta$ to adjust the duration. Here, we highlight a feature, namely the asymmetry between expansion and compression. Figure~\ref{fig:S_energy}(b) compares the two processes in the quasi‑isothermal stroke ($\xi=0$), with compression ($u=3/2$, left) and expansion ($u=2/3$, right). In the long‑time regime both display the $1/\tau$-scaling, whereas in the short‑time regime the compression curve falls below this scaling and the expansion curve exceeds it, exposing a finite‑time asymmetry.

We then shift our focus from the irreversibility to energetics of the polytropic steering to elucidate how work and heat are partitioned far from equilibrium. The polytropic index $\xi$ acts as a control knob, interpolating between the isothermal ($\xi=0$) and adiabatic ($\xi=-1$) limits~\cite{ma2023}. Starting from the ensemble-level first law of thermodynamics, $d \langle E \rangle = d W + d Q$, with $\langle E \rangle = \chi \theta(t)$, we find the total work input $W = \chi \xi^{-1}T_s (1+\delta)(1-u^{-\xi})$~\cite{martinez2015,shiraishi2023} and the absorbed heat $Q = -(1+\xi)W$. As illustrated in Fig.~\ref{fig:S_energy}(c), varying $\tilde{\tau}$ provides quantitative control over the magnitude of energy transfer. The isothermal and adiabatic cases emerge as special points where the injected work is either entirely dissipated as heat ($Q+W=0$, $\Delta E=0$) or entirely stored/released as internal energy ($Q=0$), respectively. This unified description highlights how finite-time driving can simultaneously modulate energy redistribution and the accompanying thermodynamic cost. 

\begin{figure}
	\centering
	\includegraphics[width=0.49\textwidth]{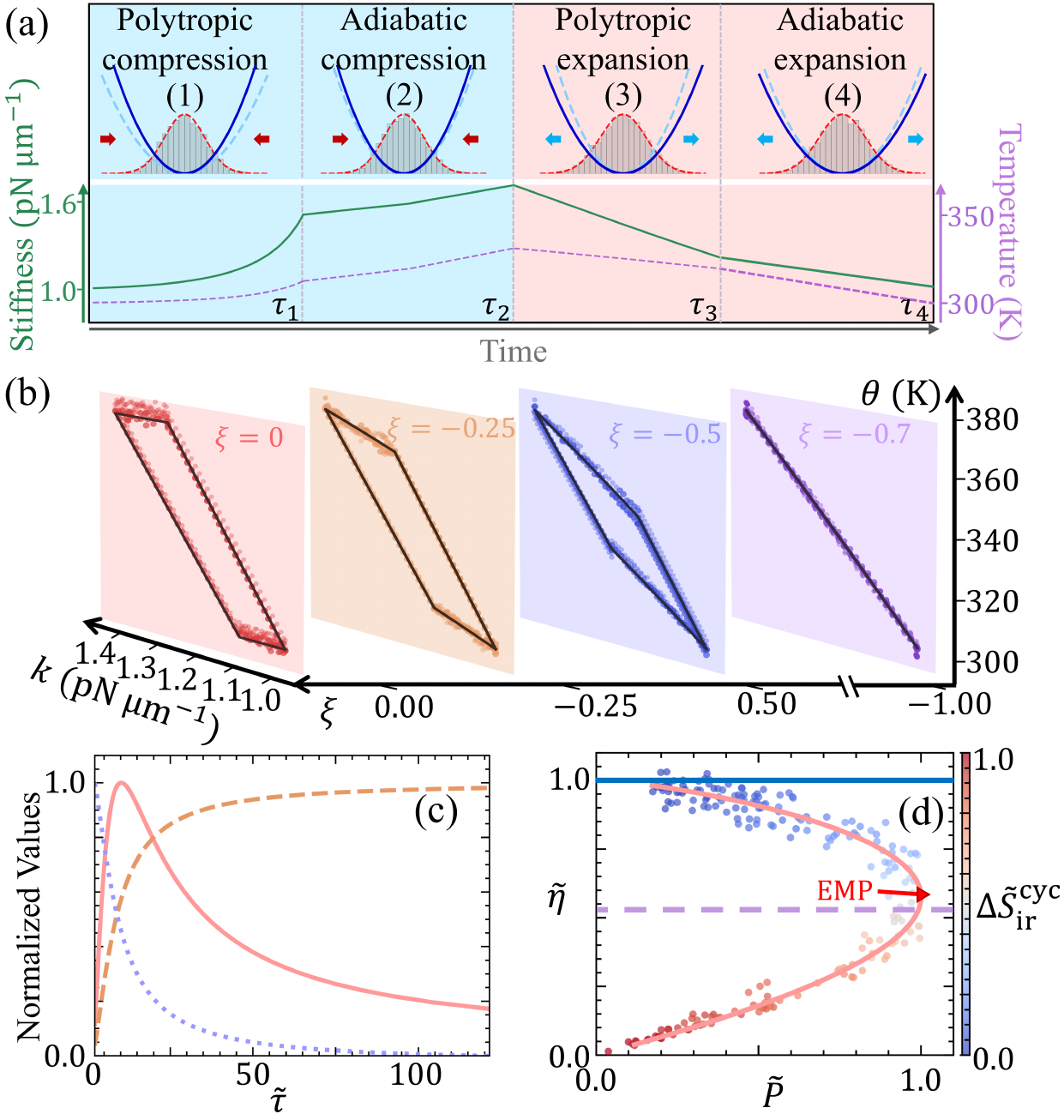}
	\caption{(a) Time evolution of the cycle protocol. (b) $k-\theta$ diagrams of the cycles with different $\xi$. (c) Normalized power $\tilde{P} \equiv P/P_{\mathrm{max}}$ (pink solid curve), efficiency $\eta \equiv \eta/\eta_{\mathrm{C}}$ (orange dashed curve), and $\Delta \tilde{S}^{\mathrm{cyc}}_{\mathrm{ir}}\equiv \Delta S^{\mathrm{cyc}}_{\mathrm{ir}}/\mathrm{Max}[\Delta S^{\mathrm{cyc}}_{\mathrm{ir}}]$ (blue dotted curve) versus $\tilde{\tau}$. (d) The power–efficiency trade-off (pink solid curve). Simulation data points are color‑coded by $\Delta \tilde{S}^{\mathrm{cyc}}_{\mathrm{ir}}$ using the coolwarm colormap and the blue solid line, red triangle marker and purple dashed line indicate the Carnot efficiency, the EMP of polytropic cycle and CA efficiency, respectively.}
	\label{fig:cycle}
\end{figure}

\textit{Brownian polytropic cycles}.--- The experimental realization of thermodynamic cycles for Brownian particles is a formidable challenge, primarily due to the technical difficulty of simultaneously modulating the confinement potential and the particle's effective temperature along a prescribed path \cite{martinez2016}. The current proposed polytropic steering framework resolves this by providing a unified, precisely controllable protocol that spans a broad class of thermodynamic processes. Once initialized, the polytropic constraint ensures that the effective temperature follows a deterministic trajectory, so the experimentalist only needs to modulate the trap stiffness via Eq.~\eqref{eq:over_general_proto} without actively controlling the temperature. To demonstrate its versatility, we construct a generic four-stroke heat engine cycle consisting of two polytropic strokes ($\xi \in [-1,0]$) interleaved by two rapid adiabatic transitions ($\xi=-1$). As shown in the upper panel of Fig.~\ref{fig:cycle}(a), the probability distributions obtained from Langevin simulations (gray histograms) perfectly match our theoretical predictions (red dashed lines), validating the stability of the steering under thermal fluctuations~\cite{foot1}. The cycle operates as follows: (i) polytropic compression while rejecting heat to a cold reservoir ($T_c$); (ii) adiabatic compression; (iii) polytropic expansion while absorbing heat from a hot reservoir ($T_h$); and (iv) adiabatic expansion back to the initial state. Crucially, as exemplified for the case of $\xi=-0.2$ in Fig.~\ref{fig:cycle}(a) (lower panel), the time-dependent trap stiffness $k_t$ and the effective temperature $\theta$ are co-modulated in a deterministic manner, ensuring the system strictly adheres to the polytropic invariant throughout the finite-time operation.

By integrating the overdamped Langevin dynamics with our derived protocols, we characterize the engine's performance across the full parametric range. Fig.~\ref{fig:cycle}(b) maps these cycles in the $k-\theta$ (stiffness-temperature) state space. The remarkable agreement between simulation (markers) and theory (solid lines) underscores the robustness of the framework. We observe a clear geometric transition: at $\xi=0$, the cycle operates as an endoreversible Carnot-like cycle; as $\xi$ decreases toward $-1$, the cycle area, representing the net work performed per cycle, gradually contracts, eventually collapsing into a single adiabatic curve where the work-heat exchange vanishes. Here, $\xi$ emerges as a thermodynamic control knob that not only dictates the protocol shape but also governs the energy partitioning between work and heat of the cycles, offering a high degree of control over the engine's output. The performance of heat engines is quantified by the efficiency $\eta=-W_{\mathrm{net}}/Q_{\mathrm{in}}$ and average output power $P=-W_{\mathrm{net}}/\tau_{\mathrm{cyc}}$. Here, $W_{\mathrm{net}}=\sum W_i~(i=1,2,3,4)$, while the heat input $Q_{\mathrm{in}}$ is defined by the energy absorbed from the hot reservoir, and $\tau_{\mathrm{cyc}}=\sum \tau_i$ is the total cycle duration. 

The practical power of this framework lies in its ability to map the complete power-efficiency relation of microscopic engines. In Fig.~\ref{fig:cycle}(c), the time evolution of the normalized power $\tilde{P}$, efficiency $\tilde{\eta}$, and cycle irreversible entropy generation (IEG, $\Delta\tilde{S}_{\rm ir}^{\rm cyc}$) versus the cycle duration $\tilde{\tau}$ clearly reveals the underlying thermodynamic constraints. Notably, the cycle IEG exhibits a monotonic decrease as $\tilde{\tau}$ increases~\cite{foot3}. Since the cycle IEG quantifies the dissipation, its monotonic decay is consistent with the monotonic rise of the efficiency $\tilde{\eta}$, which approaches the Carnot limit ($\tilde{\eta} \to 1$). In contrast, the power $\tilde{P}$ exhibits a non-monotonic dependence on $\tilde{\tau}$: it is suppressed at the fast-driving limit ($\tilde{\tau} \to 0$) due to zero work, peaks at an intermediate duration representing the optimal compromise, and vanishes in the quasistatic limit ($\tilde{\tau} \to \infty$) due to the infinitely long cycle time.

The phase diagram in Fig.~\ref{fig:cycle}(d), obtained by eliminating time from Fig.~\ref{fig:cycle}(c), visualizes how efficiency varies with power across different protocols. The theoretical prediction (pink solid curve) shows excellent agreement with numerical simulations. By color-coding the simulation points with the normalized cycle IEG, the physical picture becomes clear: the thermodynamic cost governs the accessible operating regimes. The fast-driving processes (red) confine the engine to the low-efficiency region, while a reduction in IEG (transitioning to blue) steers the operating point through the regime of maximum power and ultimately toward the near-reversible Carnot limit. The efficiency at maximum power (EMP) of the polytropic cycle, highlighted by the red triangle marker, exceeds the Curzon-Ahlborn (CA) efficiency $\eta_{\mathrm{CA}}=1-\sqrt{T_c/T_h}$ (purple dashed line) for the parameters considered. Furthermore, its dependence on $\delta$ is non-monotonic, as demonstrated in the ideal-gas heat engines in the Supplemental Material~\cite{foot1}. The framework thus offers a practical route for rational design of high-performance microscopic heat engines. By varying $\xi$ (equivalently, $\tau$), the framework scans the entire parameter space from the isothermal to the adiabatic limit, yielding the complete power-efficiency relation that identifies the optimal operating regimes and their associated dissipative costs.

\textit{Discussion}.--- Our findings challenge the conventional perception \textit{faster requires more consumption} for finite-time processes in non-equilibrium thermodynamics. By introducing the finite-time polytropic steering, we demonstrate that irreversibility is not an inevitable monotonic penalty for faster driving. Instead, the emergence of a 'most-irreversible' timescale signifies a unique dissipative peak, beyond which rapid driving actually suppresses entropy generation. Beyond its theoretical elegance, the polytropic index $\xi$ emerges as a thermodynamic control knob, enabling the precision engineering of thermodynamic trajectories. By co-modulating control parameters along prescribed polytropic paths, this paradigm allows experimentalists to navigate and optimize the complex power-efficiency relation. Furthermore, its finite-time polytropic steering conceptual structure can be extended to other systems. As a concrete illustration, we provide in the Supplemental Material a detailed analysis of polytropic steering for an ideal gas~\cite{foot1}, showing that the same protocols and the non-monotonic irreversibility persist in this macroscopic setting. This extensibility is of practical significance for engineering applications, as many real-world thermal machines operate along polytropic-like pathways where heat exchange and work extraction are coupled in finite time~\cite{dai2019}. 

Looking forward, the polytropic steering paradigm opens several avenues for exploration. First, the extension of this framework into the quantum regime promises to unveil how quantum coherence and entanglement modulate the non-monotonic irreversibility, potentially offering a quantum speedup for thermodynamic control~\cite{villazon2019} and possible synergies between polytropic steering and Mpemba-like effects for accelerating finite-time thermodynamic processes would be an interesting direction for future investigation~\cite{Lin2022,Joshi2024,Moroder2024,Liu2024,Strachan2025,Van2025,Nava2025,Li2025}. Second, in the realm of active matter, where systems are driven by non-conservative internal forces, our steering protocol could provide a systematic route to harness energy from active baths, redefining the boundaries of work extraction in non-reciprocal environments~\cite{davis2024,Wang2025}. Finally, by integrating our approach with information thermodynamics, one could explore the finite-time costs of information erasure and feedback control, potentially pushing the limits of high-speed, low-dissipation information processing and computing~\cite{proesmans2020}. Our work establishes a versatile blueprint for the next generation of high-speed, high-efficiency energy converters from the sub-atomic to the industrial scale.

\textit{Acknowledgements}.--- We thank J.-C.C. for helpful comments. Y.-H.M. thanks the National Natural Science Foundation of China for support under grant No. 12305037 and the Fundamental Research Funds for the Central Universities under grant No. 2023NTST017. S.H.S thanks the National Natural Science Foundation of China for support under Grant No.12675048 and 12364008, the Natural Science Foundation of Fujian Province for support under Grant No.2023J01006, and the Fundamental Research Fund for the Central Universities of China for support under Grant No.20720240145.

\bibliography{main}

\newpage

\onecolumngrid{}

\begin{center}
\textbf{End Matter}
\end{center}

\twocolumngrid{}

\begin{figure}
	\centering
\includegraphics[width=0.48\textwidth]{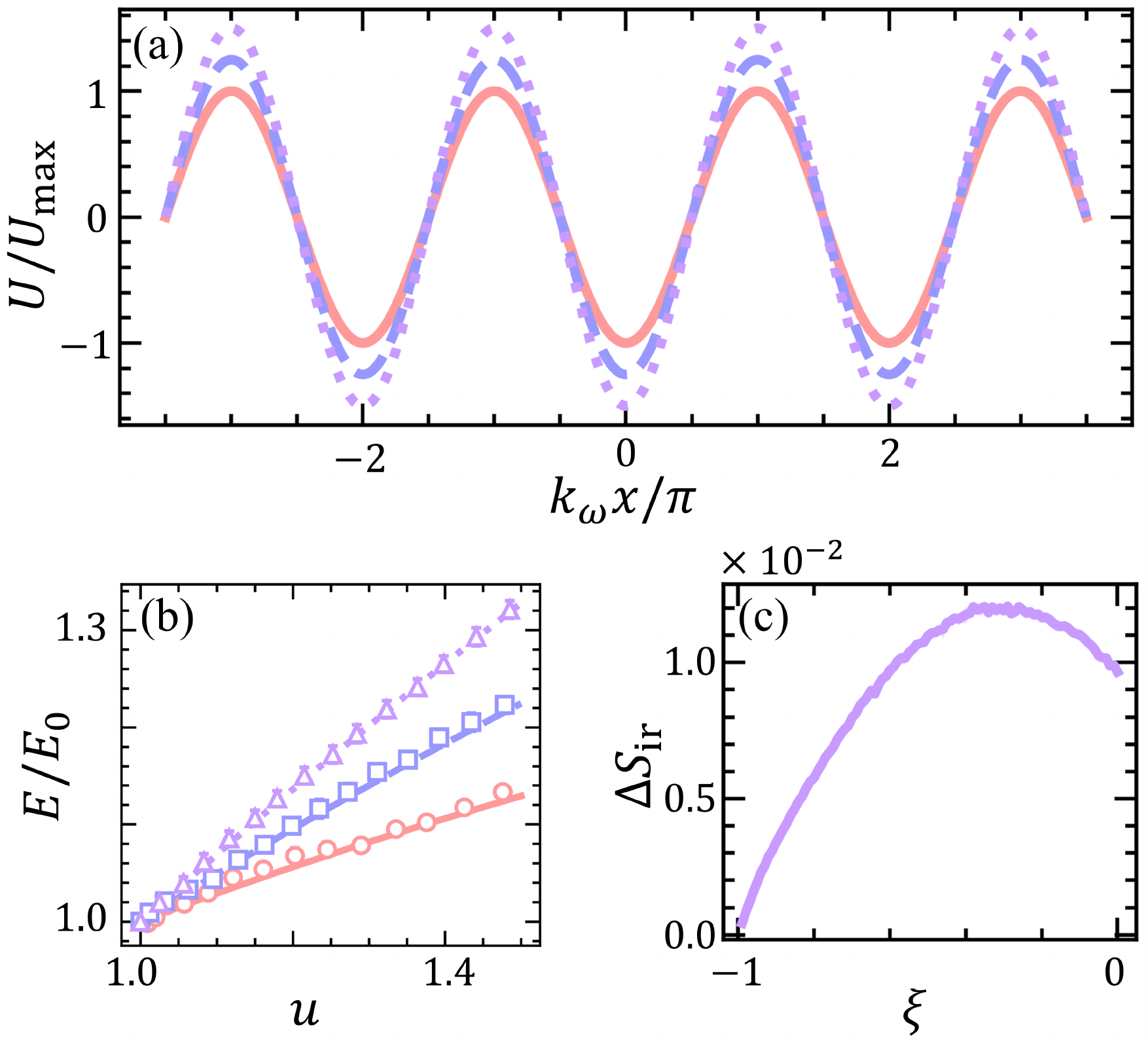}
	\caption{(a) Optical lattice potential $U(x)/U_{\mathrm{max}}$ for $\Omega_0$ (pink solid), $1.25\Omega_0$ (blue dashed), and $1.5\Omega_0$ (purple dotted). (b) Normalized energy $E/E_0$ versus $u$ for $\xi=-0.3$, $-0.5$, and $-0.7$; symbols denote Langevin simulation results and curves show theoretical predictions. (c) IEG $\Delta S_{\mathrm{ir}}$ versus $\xi$ for the optical lattice potential.}
	\label{fig:optical}
\end{figure}

The general framework developed in this Letter is not restricted to harmonic confinement. As a representative anharmonic system, we consider the optical lattice potential
\begin{equation}
	U(x,\Omega(t))=\Omega(t)[ \frac{1}{2}(k_\omega x)^2 -\alpha \cos (k_\omega x) ],
	\label{eq:optical lattice potential}
\end{equation}
where $k_\omega$ is the wave number of the optical lattice, $\alpha >0$ characterizes the lattice depth, and $\Omega(t)$ is the time-dependent control parameter that modulates the overall stiffness, as demonstrated in Fig.~\ref{fig:optical}(a). This potential combines harmonic confinement with a periodic modulation and is widely used in cold-atom and optical-tweezer experiments~\cite{Jaksch1998,lacki2021}. Unlike the harmonic trap, no exact Gaussian solution exists for the corresponding Fokker-Planck equation, so neither the steering protocol nor the irreversible entropy generation admits a closed analytical expression.

Nevertheless, the general framework remains directly applicable. To apply Eq.~\eqref{eq:over_general_proto}, it is convenient to introduce
\begin{equation}
V(x)
\equiv
\frac{1}{2}(k_\omega x)^2
-\alpha\cos(k_\omega x),
\qquad
U(x,\Omega)=\Omega V(x).
\end{equation}
The required derivatives are
\begin{equation}
\partial_x U
=
\Omega
\left[
k_\omega^2x
+
\alpha k_\omega\sin(k_\omega x)
\right],
\label{eq:optical_Ux}
\end{equation}
\begin{equation}
\partial_{xx}U
=
\Omega k_\omega^2
\left[
1+\alpha\cos(k_\omega x)
\right],
\label{eq:optical_Uxx}
\end{equation}
and
\begin{equation}
\partial_\Omega U
=
V(x)
=
\frac{U}{\Omega}.
\label{eq:optical_Ulambda}
\end{equation} 
Consequently, the ensemble-averaged energy and the moments entering Eq.~\eqref{eq:over_general_proto} become
\begin{equation}
E(t)
=
\Omega
\int_{-\infty}^{\infty}
V(x)\rho(x,t)\,dx,
\label{eq:optical_energy}
\end{equation}
\begin{equation}
\left\langle(\partial_xU)^2\right\rangle
=
\Omega^2
\int_{-\infty}^{\infty}
\left[
k_\omega^2x
+
\alpha k_\omega\sin(k_\omega x)
\right]^2
\rho(x,t)\,dx,
\label{eq:optical_force_square}
\end{equation}
and
\begin{equation}
\left\langle\partial_{xx}U\right\rangle
=
\Omega k_\omega^2
\int_{-\infty}^{\infty}
\left[
1+\alpha\cos(k_\omega x)
\right]
\rho(x,t)\,dx.
\label{eq:optical_curvature}
\end{equation}
Because $U$ is linear in $\Omega$,
\begin{equation}
\Omega\left\langle\partial_\Omega U\right\rangle
=
\langle U\rangle
=
E.
\label{eq:optical_scaling_identity}
\end{equation}
Substituting into Eq.~\eqref{eq:over_general_proto} therefore gives the optical-lattice control equation
\begin{equation}
\begin{split}
    \dot{\Omega}
&=
\frac{\Omega^2}
{m\gamma(1+\xi)E}
\left\{
\Omega
\left\langle
\left[
k_\omega^2x
+
\alpha k_\omega\sin(k_\omega x)
\right]^2
\right\rangle \right.\\
&\left. -
T_{\rm s}k_\omega^2
\left\langle
1+\alpha\cos(k_\omega x)
\right\rangle
\right\}.
\label{eq:optical_control}
\end{split}
\end{equation}

Equation~\eqref{eq:optical_control}, together with the full Fokker--Planck equation,
\begin{equation}
\partial_t\rho
=
\frac{\partial}{\partial x}
\left[
\frac{1}{m\gamma}
\left(
\Omega
\left[
k_\omega^2x
+
\alpha k_\omega\sin(k_\omega x)
\right]\rho
+
T_{\rm s}\partial_x\rho
\right)
\right],
\label{eq:optical_FP}
\end{equation}
forms a closed integro-differential control problem for $\rho(x,t)$ and $\Omega(t)$. Importantly, all averages in Eq.~\eqref{eq:optical_control} are evaluated over the evolving nonequilibrium distribution rather than an assumed equilibrium. Thus, the anharmonic deformation of the distribution is retained.

For a prescribed polytropic index $\xi$, the steering condition requires
\begin{equation}
E(t)\Omega^\xi(t)
=
E_0\Omega_0^\xi.
\end{equation}
Numerically integrating Eqs.~\eqref{eq:optical_control} and~\eqref{eq:optical_FP} yields the control protocol and the corresponding nonequilibrium distribution. Figure~\ref{fig:optical}(b) compares the resulting energy trajectories for three representative polytropic indices ($\xi=-0.3,~-0.5,~-0.7$). The numerical solutions (solid curves) of Eq.~\eqref{eq:over_general_proto} show good agreement with direct Langevin simulations (symbols). The close agreement demonstrates that the prescribed polytropic relation is accurately maintained even in the presence of the periodic anharmonic modulation. This provides a direct realization of finite-time polytropic steering beyond the analytically solvable harmonic trap.

The same nonequilibrium distribution can be used directly in the general entropy generation formula, Eq.~\eqref{general_IEG_over_2}. For the optical-lattice potential, its three contributions become
\begin{align}
\dot{\Sigma}(t)
={}&
\frac{\Omega^2}{m\gamma T_{\rm s}}
\left\langle
\left[
k_\omega^2x
+
\alpha k_\omega\sin(k_\omega x)
\right]^2
\right\rangle
\nonumber\\
&-
\frac{2\Omega k_\omega^2}{m\gamma}
\left\langle
1+\alpha\cos(k_\omega x)
\right\rangle
+
\frac{T_{\rm s}}{m\gamma}
\int_{-\infty}^{\infty}
\frac{[\partial_x\rho(x,t)]^2}
{\rho(x,t)}
\,dx .
\label{eq:optical_entropy_rate}
\end{align}
Thus, the total irreversible entropy generation is
\begin{equation}
    \Delta S_{\rm ir}
=
\int_0^\tau
\dot{\Sigma}(t)\,dt .
\label{eq:optical_entropy_total}
\end{equation}
This expression requires the full nonequilibrium distribution and therefore has no general closed form for the optical lattice, in contrast to the harmonic potential, for which Eq.~\eqref{general_IEG_over_2} can be integrated analytically.

By utilizing the integration-by-parts identity,
\begin{equation}
\int_{-\infty}^{\infty}
V'(x)\partial_x\rho(x,t)\,dx
=
-
\int_{-\infty}^{\infty}
V''(x)\rho(x,t)\,dx,
\end{equation}
Eq.~\eqref{eq:optical_entropy_rate} can be equivalently written as
\begin{equation}
    \dot{\Sigma}(t)
=
\frac{1}{m\gamma T_{\rm s}}
\int_{-\infty}^{\infty}
\rho(x,t)
\left[
\Omega V'(x)
+
T_{\rm s}\partial_x\ln\rho(x,t)
\right]^2
dx
\geq 0.
\label{eq:optical_entropy_positive}
\end{equation}
Hence, the positivity of the irreversible entropy generation is independent of the harmonic approximation and follows directly from the full Fokker-Planck dynamics.

Figure~\ref{fig:optical}(c) shows the total entropy generation obtained from the corresponding Langevin simulations. Despite the absence of an analytical expression, $\Delta S_{\rm ir}$ retains the characteristic non-monotonic dependence on the polytropic index: the dissipation increases from the fast-driving side, reaches a maximum at an intermediate $\xi$, and decreases again toward the slow-driving limit. This persistence demonstrates that the non-monotonic irreversibility is not a peculiarity of Gaussian harmonic dynamics.

The two limiting regimes also acquire an interpretation within the general framework. For $\xi\rightarrow0$, the polytropic constraint becomes
\begin{equation}
E(t)=E_0,
\end{equation}
corresponding to the endoreversible branch of the construction. In the harmonic, locally equilibrated limit, this reduces to the isothermal condition because $E=\theta/2$. For the general anharmonic system, however, $E$ should be regarded as the controlled energetic variable rather than identified with an instantaneous effective temperature.

Conversely, in the adiabatic limit $\xi\rightarrow-1$, the
polytropic constraint gives
\begin{equation}
\frac{E(t)}{\Omega(t)}
=
\frac{E_0}{\Omega_0}.
\end{equation}
Hence,
\begin{equation}
\dot{E}
=
\frac{E}{\Omega}\dot{\Omega}.
\end{equation}
For the optical-lattice potential $U=\Omega V$, one has
\begin{equation}
\left\langle\partial_\Omega U\right\rangle
=
\frac{\langle U\rangle}{\Omega}
=
\frac{E}{\Omega}.
\end{equation}
It therefore follows that
\begin{equation}
\dot{E}
=
\dot{\Omega}
\left\langle\partial_\Omega U\right\rangle
=
\dot{W},
\end{equation}
and consequently
\begin{equation}
\dot{Q}=\dot{E}-\dot{W}=0.
\end{equation}
Thus, the $\xi\rightarrow-1$ limit corresponds to an adiabatic
process with vanishing heat transfer, independently of the
anharmonicity of the confining potential. The two limiting branches therefore suppress accumulated irreversibility through distinct mechanisms, the reservoir-equilibrated quasistatic isothermal limit at $\xi\rightarrow0$ and the vanishing-heat-transfer adiabatic limit at $\xi\rightarrow-1$, while intermediate polytropic indices produce the competition between driving and thermal relaxation.

These results establish that harmonic confinement is not the origin of either the finite-time polytropic construction or its non-monotonic irreversibility. Rather, the harmonic trap is distinguished by analytical tractability, whereas Eqs.~\eqref{eq:over_general_proto} and~\eqref{general_IEG_over_2} provide a general framework that remains applicable to anharmonic potentials through the full nonequilibrium distribution. The optical-lattice realization therefore demonstrates the predictive capability of the framework beyond Gaussian dynamics and establishes finite-time polytropic steering as a general construction for nonequilibrium Fokker--Planck systems.

\clearpage
\onecolumngrid{}
\mbox{} 
\begingroup
\pagestyle{empty}

\includepdf[
pages=-,
pagecommand={},
turn=false
]{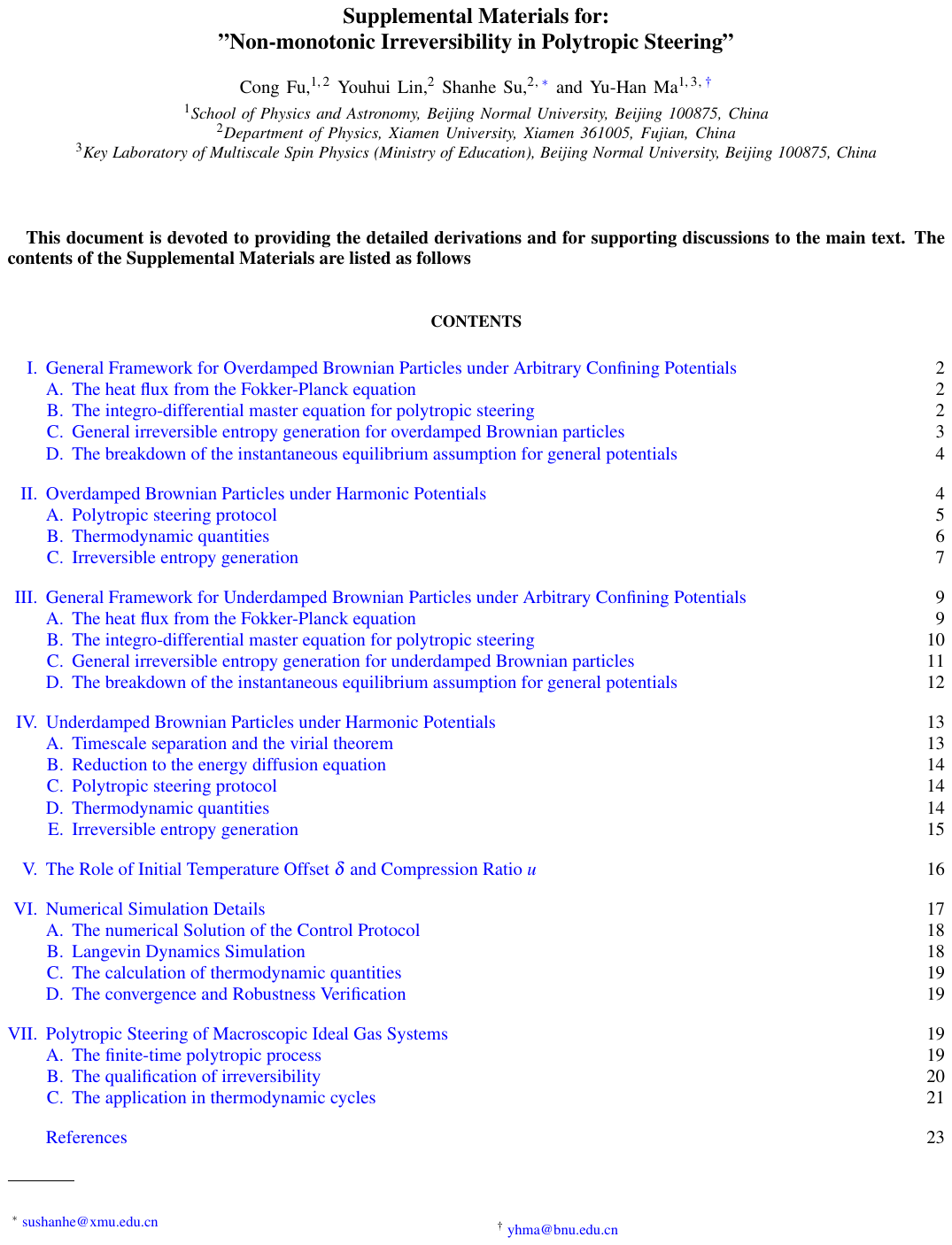}

\endgroup
\clearpage

\end{document}